\documentclass[aps,pra,twocolumn]{revtex4}

\begin{document}
\title{Optimal estimation of a physical observable's expectation value 
for pure states}
\author{A. Hayashi, M. Horibe, and T. Hashimoto}
\affiliation{Department of Applied Physics\\
           University of Fukui, Fukui 910-8507, Japan}

\begin{abstract}
We study the optimal way to estimate the quantum expectation value of a 
physical observable when a finite number of copies of a quantum pure 
state are presented. The optimal estimation is determined by minimizing the 
squared error averaged over all pure states distributed in a unitary invariant 
way. We find that the optimal estimation is ``biased'' though the optimal 
measurement is given by successive projective measurements of the 
observable. The optimal estimate is not the sample average of observed data, 
but the arithmetic average of observed and ``default nonobserved'' data, 
with the latter consisting of all eigenvalues of the observable.
\end{abstract}

\maketitle

\newcommand{\ket}[1]{|\,#1\,\rangle}
\newcommand{\bra}[1]{\langle\,#1\,|}
\newcommand{\braket}[2]{\langle\,#1\,|\,#2\,\rangle}
\newcommand{\bold}[1]{\mbox{\boldmath $#1$}}
\newcommand{\sbold}[1]{\mbox{\boldmath ${\scriptstyle #1}$}}
\newcommand{\tr}[1]{{\rm tr}\!\left[#1\right]}
\newcommand{\trm}{{\rm tr}}
\newcommand{\BC}{{\bold{C}}}
\newcommand{\CS}{{\cal S}}
\newcommand{\CH}{{\cal H}}

\section{Introduction}
One of the fundamental tasks in quantum physics is to determine the 
expectation value of a physical observable of an unknown quantum 
state. With only a single copy of the quantum state given, we cannot determine 
the expectation value of a physical observable because of the statistical 
nature of quantum measurement. 
Suppose we are presented with a certain finite number 
$N$ of copies of an unknown quantum state. We cannot increase the number of 
the copies since the no-cloning theorem \cite{Wootters82} forbids it. 
Then what is the optimal way to determine the expectation value of the 
observable for a given $N$? 
An intuitively plausible optimal estimate is given by the arithmetic average 
of the data produced by successive projective measurements of the 
observable on the individual systems. 

This problem, however, is by no means trivial. 
Given a quantum system composed of subsystems, we can consider two types 
of measurement. One is separate measurements: a sequence of measurements 
on the individual subsystems, possibly dependent on the outcomes of earlier 
measurements. The other is joint measurement: a single measurement on the 
system as a whole.
Recent studies on quantum-state discrimination and 
estimation \cite{Helstrom76, Holevo82} provide considerable instances in 
which joint measurements perform better than separate measurements even 
for a state composed of mutually uncorrelated subsystems. 

Peres and Wootters showed that a certain set of three bipartite product 
states can be better distinguished by a joint measurement 
\cite{Peres91,Wootters05} (see also \cite{Ban97,Sasaki98,Eldar01}). 
An even stronger example was provided by Bennett {\it et al.} 
\cite{Bennett99}, which shows that a certain orthogonal set of bipartite 
product states cannot be reliably distinguished by any separate measurement 
though a joint measurement perfectly distinguishes them because of their 
mutual orthogonality. 
The superiority of joint measurement has also been discussed in the problem of 
quantum-state estimation for identically prepared copies of an unknown state 
(see \cite{Massar95,Derka98,Bagan01,MHayashi98,Holevo03,Hayashi05_estimation}, 
for example).

In \cite{D'Ariano05}, D'Ariano, Giovannetti, and Perinotti raised the question 
of whether the standard procedure of averaging the outcomes of repeated 
measurements of an observable over equally prepared systems is the best way 
of estimating the expectation value of the observable, or whether a joint 
measurement 
can improve the estimation. They showed that the standard procedure is indeed 
optimal if one is restricted to the class of unbiased estimation for 
any generally mixed state. Here an estimator is said to be unbiased if 
the average over many independent estimates gives the true value to be 
estimated.

An unbiased result is certainly one of the desirable properties for estimation 
but not a necessary condition. 
A natural question is then whether a ``biased'' estimation performs 
better than the standard unbiased estimation. 
Let us take a simple example, in which we estimate the expectation 
value of the observable $\sigma_z$ for a single-qubit system in an unknown 
pure state. We assume that the state of the qubit is chosen according to 
the uniform distribution on the Bloch sphere. Suppose that the projective 
measurement of $\sigma_z$ produced the outcome $1$, which means the sample 
average is 1. 
Now, one can ask if it is reasonable to conclude that the expectation value 
of $\sigma_z$ is most likely equal to 1. 
Note that the expectation value of $\sigma_z$ is 1 only if the qubit lies 
exactly at the north pole of the Bloch sphere. On the other hand, the 
measurement of $\sigma_z$ can produce the outcome 1 with some probability 
unless the qubit is exactly at the south pole. 
Therefore, it is more reasonable to consider that the expectation value of 
$\sigma_z$ is not 1, but somewhere between 0 and 1. 
In fact, the optimal estimate turns out to be 1/3 in this case, as we will see 
in the next section.

In this paper, without assuming unbiasedness of the estimation, we study 
the optimal procedure for the expectation value of a physical observable of 
an unknown pure state, when $N$ copies of the state are presented. 
We assume that the unknown pure state is chosen from the pure-state space 
according to a unitary invariant {\it a priori} distribution. 
The optimal estimation is determined by minimizing the squared error averaged 
over the {\it a priori} distribution.

\section{Optimal estimation}
We determine the optimal way to estimate the expectation value of 
a physical observable $\Omega$ when $N$ copies of an unknown pure state 
$\rho=\ket{\phi}\bra{\phi}$ on a $d$-dimensional Hilbert space  
$\CH$ are given. Let $\{E_a\}$ be a positive-operator-valued measure 
(POVM) on the total system $\CH^{\otimes N}$, with outcome labeled 
$a$ providing an estimate $\omega_a$ for the expectation value 
given by $\tr{\rho\Omega}$. For a given $\rho$, the mean squared error 
in the estimate is written as 
\begin{eqnarray}
  \Delta(\rho) = \sum_a \tr{E_a\rho^{\otimes N}}
             \left( \omega_a - \tr{\rho\Omega} \right)^2.
                     \label{Delta_rho}
\end{eqnarray}
We will first average this $\Delta(\rho)$ over all pure states $\rho$ and 
then minimize it with respect to the POVM $\{E_a\}$ and the estimate 
$\{\omega_a\}$. 

The distribution of the pure states $\rho$ is specified 
in the following way. Expand a pure state as 
$\ket{\phi}=\sum_{i=1}^d c_i\ket{i}$ in terms of an 
orthonormal base $\{\ket{i}\}$ of $\CH$. The distribution is then defined to 
be the one in which the $2d$-component real vector 
$\{x_i={\rm Re}\,c_i, y_i={\rm Im}\,c_i \}$ is uniformly distributed on the 
$(2d-1)$-dimensional hypersphere of radius 1. 
The distribution is unitary invariant in the sense that it is independent of 
the orthonormal base $\{\ket{i}\}$ chosen to define it. 
Let us denote the average over this distribution by $\langle \cdots \rangle$. 
All we need in the following calculation is a useful relation for the 
average of $\rho^{\otimes n}$ given in Ref.~\cite{Hayashi05_estimation}, 
that is, 
\begin{eqnarray}
   \langle \rho^{\otimes n} \rangle = \frac{\CS_n}{d_n},  
                              \label{formula_1}
\end{eqnarray}
where $\CS_n$ is the projection operator onto the totally symmetric 
subspace of $\CH^{\otimes n}$ and $d_n$ is its dimension given by 
$d_n=\tr{\CS_n}={}_{n+d-1}C_{d-1}$. It may be instructive to see how 
this formula comes out in some simple cases of qubits ($d=2$), in which the 
above distribution means that the Bloch vector $\bold n$ is uniformly 
distributed on the surface of the Bloch sphere. 
Then we can easily verify 
\begin{eqnarray} 
  \frac{1}{4\pi} \int\!{\rm d}\bold{n}\, 
    \left(\frac{1+\bold{n}\cdot\bold{\sigma}}{2}\right)^{\otimes 2} 
  &=& \frac{1}{24}\left( \bold{\sigma}(1) + \bold{\sigma}(2) \right)^2
                             \nonumber \\  
  &=& \left\{ 
           \begin{array}{ll}
                 1/3   & (S=1), \\
                 0     & (S=0), \\
           \end{array}
    \right.
\end{eqnarray} 
where $S$ is the eigenvalue of the total spin. 
This is a special case of the formula (\ref{formula_1}), 
since the state is  symmetric if $S=1$ and antisymmetric if $S=0$. 
The case of three qubits provides another example.
\begin{eqnarray}
  \frac{1}{4\pi} \int\!{\rm d}\bold{n}\, 
    \left(\frac{1+\bold{n}\cdot\bold{\sigma}}{2}\right)^{\otimes 3} 
  = \left\{ 
           \begin{array}{ll}
                 1/4   & (S=3/2), \\
                 0     & (S=1/2). \\
           \end{array}
    \right.
\end{eqnarray}

Now going back to the general-dimensional case, we expand 
Eq. (\ref{Delta_rho}) and perform averaging over $\rho$ by use of the 
formula (\ref{formula_1}):
\begin{eqnarray}
   \langle \Delta \rangle 
   &=& 
       \sum_a \left\langle \tr{E_a\rho^{\otimes N}} 
          \left( \omega_a^2 -2\omega_a\tr{\rho\Omega} + 
                   (\tr{\rho\Omega})^2 \right) \right\rangle
                             \nonumber \\
   &=& \left\langle 
          \Delta_1(\rho) + \Delta_2(\rho) + \Delta_3(\rho) 
       \right\rangle,
                    \label{Delta_rho_expanded}
\end{eqnarray}
where we denote the three terms in $\Delta(\rho)$ by 
$\Delta_1(\rho)$, $\Delta_2(\rho)$, and $\Delta_3(\rho)$, 
and we evaluate each separately.
The first term $\langle \Delta_1 \rangle$ is readily 
calculated as
\begin{eqnarray}
  \langle \Delta_1 \rangle = \frac{1}{d_N} \sum_a \omega_a^2 \tr{E_a\CS_N}.
\end{eqnarray} 
For $\langle \Delta_3 \rangle$, 
we first use the completeness of the POVM by summing over $a$
and perform the average in the following way:
\begin{eqnarray}
 \langle \Delta_3 \rangle &=& \left\langle (\tr{\rho\Omega})^2 \right\rangle 
    = \left\langle \tr{\rho^{\otimes 2}\Omega(1)\Omega(2)} \right\rangle 
                                 \nonumber \\
    &=& \frac{1}{d_2} \tr{\CS_2 \Omega(1)\Omega(2)} 
      = \frac{1}{d(d+1)}\left( \left({\rm tr}\Omega\right)^2 + 
                                 \trm\Omega^2 \right),
                                 \nonumber \\
                     \label{Delta_3}
\end{eqnarray}
where $\rho^{\otimes 2}$ is understood to be the tensor product of two 
$\rho$'s in spaces 1 and 2, and the space on which the operator $\Omega$ acts 
is specified by the number in the parentheses. Hereafter we will 
use this convention in more general cases, namely, 
\begin{eqnarray}
    \Omega(n) \equiv \bold{1}^{\otimes (n-1)} \otimes \Omega
                \otimes \bold{1} \otimes \bold{1} \otimes \cdots.
\end{eqnarray}
Evaluation of the second term $\langle \Delta_2 \rangle$ is more involved. 
Introducing another system on $\CH$, which we call system $N+1$, we have 
\begin{eqnarray}
  \langle \Delta_2 \rangle &=& -2\sum_a \omega_a 
                        \left\langle \tr{E_a\rho^{\otimes N}}
                                  \tr{\rho\Omega} \right\rangle 
                               \nonumber \\
  &=& -2\sum_a \omega_a \left\langle 
            \tr{E_a\rho^{\otimes (N+1)}\Omega(N+1)} \right\rangle
                               \nonumber \\
  &=& -\frac{2}{d_{N+1}}\sum_a \omega_a 
            \tr{E_a\CS_{N+1} \Omega(N+1)},
                           \label{Delta_3_1}
\end{eqnarray}
where the traces in the second and third equations are 
understood to be over systems 1,2,\ldots,$N$, and $N+1$. 
The operator $\Omega(N+1)$ acts on system $N+1$. 
The projection operator $\CS_{N+1}$ is the sum of all permutation operators of 
$N+1$ systems divided by a factor of $(N+1)!$. Any permutation of $N+1$ 
objects is either just a permutation among the first $N$ objects or 
the product of a permutation among the first $N$ objects and the transposition 
between the $(N+1)$th object and one of the first $N$ objects. 
With this observation we find, for any operator $\Omega$, 
\begin{eqnarray}
    \trm_{N+1}\left[ \CS_{N+1}\Omega(N+1) \right] 
    &=& \frac{\CS_N}{N+1}
         \left( \trm\Omega + \sum_{n=1}^N \Omega(n) \right), 
                     \nonumber \\
                     \label{formula_2}
\end{eqnarray}
where $\trm_{N+1}$ is the trace over the $(N+1)$st system. 
We use this formula to trace out the newly introduced system $(N+1)$ 
in the expression $\Delta_2$ given by Eq. (\ref{Delta_3_1}). 
The result is given by 
\begin{eqnarray}
  \Delta_2 = -\frac{2}{d_N}\sum_a \omega_a \tr{ E_a\CS_N\hat\Omega }, 
\end{eqnarray}
where we define the symmetric one-body operator $\hat\Omega$ to be
\begin{eqnarray}
   \hat{\Omega} &\equiv& \frac{1}{N+d} \left( 
                          \trm\Omega + \sum_{n=1}^N \Omega(n) \right) .
\end{eqnarray}
Combining the three averages $\langle \Delta_1 \rangle$, 
$\langle \Delta_2 \rangle$, and $\langle \Delta_3 \rangle$, 
we obtain 
\begin{eqnarray}
  \langle \Delta \rangle 
         &=& \frac{1}{d_N} \sum_a \tr{ E_a\CS_N \left(
               \omega_a^2 -2\omega_a\hat{\Omega} \right) } 
                           \nonumber \\
         &\ & + \frac{1}{d(d+1)}\left( \left({\rm tr}\Omega\right)^2 + 
                                 \trm\Omega^2 \right). 
\end{eqnarray}
To minimize $\langle \Delta \rangle$ we complete the square with respect to 
$\omega_a$ in this expression.  Owing to the completeness of the POVM, 
this is reduced to the calculation of $\tr{\CS_N \hat\Omega^2}$, 
which can be performed by using the following formulas: 
\begin{eqnarray}
 && \tr{\CS_N\Omega(n)} = \frac{d_N}{d}\trm\Omega,   \\
 && \tr{\CS_N\Omega(n)\Omega(m)}   
                          \nonumber \\
 &&\ \ \ \ = 
       \left\{
           \begin{array}{l}
              \frac{d_N}{d}\tr{\Omega^2}\ (n=m), \\
              \frac{d_N}{d(d+1)}
                 \left( \tr{\Omega^2} + (\trm\Omega)^2 \right)\ (n \ne m). \\
           \end{array}
       \right.
\end{eqnarray}
After some calculation we find 
\begin{eqnarray}
  \tr{\CS_N \hat\Omega^2}
       &=& \frac{d_N}{d(d+1)(N+d)} \times 
                          \nonumber \\
       && \left( N \tr{\Omega^2} + (N+d+1)(\trm\Omega)^2 \right). 
\end{eqnarray}
We thus finally obtain the mean squared error in the completed square form 
\begin{eqnarray}
  \langle \Delta \rangle 
   &=& \frac{1}{d_N}\sum_a 
        \tr{ E_a \CS_N  \left( \omega_a - \hat\Omega \right)^2 } 
                                     \nonumber \\
   && + \frac{1}{d(d+1)(N+d)} 
        \left( d \tr{\Omega^2} - \left( \trm\Omega \right)^2 \right).
                    \label{Delta_square}
\end{eqnarray}

Now note that the first term in Eq.~(\ref{Delta_square}) is positive. 
This is because $\hat\Omega$ is symmetric under exchange of component 
subsystems and therefore 
$\CS_N(\omega_a - \hat\Omega)^2=\CS_N(\omega_a - \hat\Omega)^2\CS_N$ 
is a positive operator. The $\Delta$ has a lower bound given by 
the second term of Eq.~(\ref{Delta_square}). 
Let us denote the eigenvalue of $\Omega$ by $\Omega_i\ (i=1,\ldots,d)$ and the 
corresponding eigenstate by $\ket{i}$. 
It is then readily seen that this lower bound can be achieved if 
the index $a$ of the POVM element collectively represents the set of 
$\{i_1,i_2,\ldots,i_N\}$,  the POVM element is taken to be the projector 
\begin{eqnarray}
       E_{i_1,i_2,\ldots,i_N} = 
           \ket{i_1i_2 \cdots i_N}\bra{i_1i_2 \cdots i_N}, 
                     \label{OmegaPOVM}
\end{eqnarray}
and the estimate $\omega_a$ to be the corresponding eigenvalue of $\hat\Omega$,
\begin{eqnarray}
    \omega_{i_1,i_2,\ldots,i_N}=\frac{1}{N+d}
          \left( \trm\Omega+\sum_{n=1}^N \Omega_{i_n} \right).
\end{eqnarray}

Thus we conclude that the mean squared error $\langle \Delta \rangle$ in the 
estimation for the expectation value of the observable $\Omega$ takes 
its minimum value given by 
\begin{eqnarray}
   \Delta_{{\rm opt}} = \frac{1}{d(d+1)(N+d)} 
        \left( d \tr{\Omega^2} - \left( \trm\Omega \right)^2 \right),
                      \label{Delta_min}
\end{eqnarray}
if one measures the observable $\Omega$ independently for each system and 
makes the estimate given by
\begin{eqnarray}
    \omega_{{\rm opt}} \equiv 
         \frac{1}{N+d} \left( \trm\Omega+\sum_{n=1}^N \Omega_{i_n} \right), 
                          \label{omega_opt}
\end{eqnarray}
where $\{\Omega_{i_1},\Omega_{i_2},\ldots,\Omega_{i_N}\}$ are the 
data observed by the measurement. 

The optimal estimate $\omega_{{\rm opt}}$ is not the arithmetic average of 
observed data (the sample average), though the optimal measurement is 
projective and independent. For a finite $N$, it is not unbiased either since 
\begin{eqnarray}
  &&   \sum_a \omega_a \tr{E_a \rho^{\otimes N}} 
                        \nonumber \\
  && = \tr{\hat\Omega\rho^{\otimes N}} 
         = \frac{1}{N+d}\left( \trm\Omega+N\tr{\rho\Omega}\right ),
\end{eqnarray}
which only asymptotically approaches $\tr{\rho\Omega}$. 
In Sec.~IV we will discuss the biasedness of $\omega_{{\rm opt}}$ and 
present an interpretation of its structure. 

What do we obtain for the mean squared error if we take the  
sample average of the values of $\Omega$ observed by the successive 
measurements on each copy? 
In this case the POVM is given by Eq.~(\ref{OmegaPOVM}) and 
the estimate by 
\begin{eqnarray}
    \omega_{i_1,i_2 \ldots,i_N} = \frac{1}{N}\sum_{n=1}^N \Omega_{i_n}
                      \equiv \omega_{{\rm av}},
\end{eqnarray}
which can be easily shown to be unbiased. The squared error for 
a given $\rho$ given in Eq.~(\ref{Delta_rho}) takes the form 
\begin{eqnarray}
   \Delta_{{\rm av}}(\rho) = \frac{1}{N} 
     \left( \tr{\rho\Omega^2} - \left( \tr{\rho\Omega} \right)^2 \right).
\end{eqnarray}
After the average over $\rho$ we have 
\begin{eqnarray}
   \Delta_{{\rm av}} = \langle \Delta_{{\rm av}}(\rho) \rangle = 
      \frac{1}{d(d+1)N} \left( 
           d\trm\Omega^2 - \left( \trm\Omega \right)^2 \right), 
                      \label{Delta_av}
\end{eqnarray}
where we used Eq.~(\ref{Delta_3}).

Comparing $\Delta_{{\rm opt}}$ and $\Delta_{{\rm av}}$, 
we find that the only difference between 
them is in the factor in the denominators, $N+d$ in $\Delta_{{\rm opt}}$ 
and $N$ in $\Delta_{{\rm av}}$. 
While $\Delta_{{\rm opt}}$ is certainly less than $\Delta_{{\rm av}}$, 
both show the same asymptotics when the number of copies goes to infinity. 
The difference becomes important when the number of copies is comparable to 
the dimension of the system.

Let us examine the example discussed in Sec.~I, in which $\sigma_z$ is 
measured with the result 1 for a single qubit in an unknown pure state 
($d=2$ and $N=1$).
In this case the observed data is $\{1\}$. The estimate by the 
sample average gives $\omega_{{\rm av}}=1$ for the expectation value of 
$\sigma_z$ with the mean squared error 
$\Delta_{{\rm av}} = 2/3$, whereas the optimal estimation 
predicts $\omega_{{\rm opt}}=1/3$ with the mean squared error 
$\Delta_{{\rm opt}} = 2/9$.

\section{Estimation with the unbiasedness condition}
In Ref.~\cite{D'Ariano05}, D'Ariano, Giovannetti, and Perinotti considered 
the  estimation for the expectation of observables under the unbiasedness 
condition for any generally mixed state $\rho^{\otimes N}$ and 
showed that the optimal estimate under the constraint is given by the sample 
average obtained by the independent successive measurement of the observable 
on each copy. In this section we briefly discuss the same problem in the pure 
state case and show the same conclusion holds. 

The unbiasedness condition is written as
\begin{eqnarray}
  \sum_a \omega_a \tr{E_a \rho^{\otimes N}} = \tr{\Omega\rho}.
                           \label{ub_con}
\end{eqnarray}
Note that $\tr{\Omega\rho}$ on the right-hand side can be expressed as
\begin{eqnarray}
    \tr{\Omega\rho} &=& \tr{\hat\Omega_{{\rm av}}\rho^{\otimes N}}, \\
      \hat\Omega_{{\rm av}} &\equiv& \frac{1}{N}\sum_{n=1}^N \Omega(n).
\end{eqnarray} 
If the unbiasedness condition (\ref{ub_con}) is assumed for any generally 
mixed state $\rho$, then it can be shown \cite{D'Ariano05} that 
\begin{eqnarray}
   \sum_a \omega_a E_a = \hat\Omega_{{\rm av}} 
                           \label{ub_mixed}
\end{eqnarray}
for any permutation-invariant POVM $\{ E_a \}$.
If we require the unbiasedness condition for any 
pure state $\rho$, we can still show that the relation (\ref{ub_mixed}) 
holds in the totally symmetric subspace of $\CH^{\otimes N}$, namely, 
\begin{eqnarray}
   \CS_N \left( \sum_a \omega_a E_a  - \hat\Omega_{{\rm av}} 
                                \right) \CS_N = 0.
                           \label{ub_pure}
\end{eqnarray}
This follows from a lemma for an operator $A$ on $\CH^{\otimes N}$: 
\begin{eqnarray*}
 && \tr{A\rho^{\otimes N}}=0\ \ \ \mbox{\rm for any pure state}\ \rho, \\
 && \mbox{\rm if and only if}\ \ \ \CS_N A \CS_N=0.
\end{eqnarray*}
The ``if" part is trivial and we sketch the proof of the ``only if" part. 
We write $\ket{\phi}=\sum_{i=1}^d c_i \ket{i}$ in terms of a basis 
$\{ \ket{i} \}$ of $\CH$, where $\rho=\ket{\phi}\bra{\phi}$. 
Then we have 
\begin{eqnarray}
  \tr{A\rho^{\otimes N}}  
  &=& \sum_{i_1 \cdots i_N,j_1 \cdots j_N} 
                 c^*_{i_1} c^*_{i_2}\cdots c^*_{i_N}
                 c_{j_1} c_{j_2}\cdots c_{j_N}  
                           \nonumber \\
  && \ \ \ \ \ \ \times \bra{i_1 i_2 \cdots i_N}A\ket{j_1 j_2 \ldots j_N} 
                           \nonumber \\
  &=& \sum_{n_i,m_i} c_1^{*n_1}c_2^{*n_2} \cdots c_d^{*n_d}
                     c_1^{m_1}c_2^{m_2} \cdots c_d^{m_d} 
                           \nonumber \\
  && \ \ \ \ \ \ \times \bra{\psi_{n_1 n_2 \cdots n_d}}
                       A\ket{\psi_{m_1 m_2 \cdots m_d}}
                           \nonumber \\
  &=& 0,      \label{occupation}
\end{eqnarray}
where the summation over integers $n_i \ge 0$ and $m_i \ge 0$ should be  
taken under the conditions $\sum_i n_i = \sum_i m_i = N$, and 
the state $\ket{\psi_{n_1 n_2 \cdots n_d}}$ is the occupation-number 
representation of symmetric states (generally not normalized), 
with $n_i$ being the occupation number 
of state $i$. Equation~(\ref{occupation}) should hold for any complex 
$c_i$, implying 
$\bra{\psi_{n_1 n_2 \cdots n_d}}A\ket{\psi_{m_1 m_2 \cdots m_d}}=0$.

The difference between the two unbiased conditions (\ref{ub_mixed}) and 
(\ref{ub_pure}) is the projection operator $\CS_N$ 
in the pure-state case. This, however, does not hamper the subsequent 
argument since the support of the operator $\rho^{\otimes N}$ for pure $\rho$ 
is the totally symmetric subspace. 

We go back to the expanded form of $\Delta(\rho)$ as in 
Eq.~(\ref{Delta_rho_expanded}), but before being averaged over $\rho$. 
By using the unbiased condition (\ref{ub_pure}) we readily find 
$\Delta_2(\rho)=-2\Delta_3(\rho)$ so that we have 
\begin{eqnarray}
  \Delta(\rho) = \sum_a \omega_a^2 \tr{E_a \rho^{\otimes N}} 
                 - \left( \tr{\rho\Omega} \right)^2. 
\end{eqnarray}
It can be shown that 
\begin{eqnarray}
  \sum_a \omega_a^2 \tr{E_a \rho^{\otimes N}} \ge 
           \tr{\hat\Omega_{{\rm av}}^2 \rho^{\otimes N} }, 
\end{eqnarray}
since in the symmetric subspace we have
\begin{eqnarray}
   0 &\le& \sum_a \left( \omega_a - \hat\Omega_{{\rm av}} \right) E_a
                \left( \omega_a - \hat\Omega_{{\rm av}} \right) 
                           \nonumber \\
     &=& \sum_a \omega_a^2 E_a - \hat\Omega_{{\rm av}}^2. 
\end{eqnarray}
It is evident that the equality holds if the POVM element $E_a$ is the 
projector of the eigenstate of $\hat\Omega_{{\rm av}}$ and the estimate 
$\omega_a$ is the corresponding eigenvalue, which is the sample average 
of the observed values of $\Omega$ for each copy. Thus the minimum value 
of the squared error in the unbiased estimation is given by 
\begin{eqnarray}
  & & \tr{\hat\Omega_{{\rm av}}^2 \rho^{\otimes N}}
                      - \left( \tr{\rho\Omega} \right)^2
                              \nonumber \\
  &=& \frac{1}{N} \left(
         \tr{\rho\Omega^2} - \left( \tr{\rho\Omega} \right)^2 \right) 
   = \Delta_{{\rm av}}(\rho),  
\end{eqnarray}
which shows that the conclusion of Ref.~\cite{D'Ariano05} holds if we restrict 
ourselves to the pure-state input ensemble.  
Averaging over $\rho$ gives $\Delta_{{\rm av}}$ given in Eq.~(\ref{Delta_av}). 

\section{Discussion and Concluding Remarks}
We have seen that the optimal estimation of the expectation 
value of a physical observable is biased, though the optimal 
measurement is given by the successive projective measurement of the 
observable. The optimal estimate $\omega_{{\rm opt}}$ is not given by the 
arithmetic average of observed data. 

We can interpret the expression (\ref{omega_opt}) of the optimal estimate 
$\omega_{{\rm opt}}$ in the following way. 
First of all, we should remember that we have full knowledge on 
properties of the observable $\Omega$ including its eigenvalues. 
Otherwise we cannot perform a measurement associated with $\Omega$. 
Then what can we expect for outcomes of the $\Omega$ measurement {\it before} 
performing the measurement? The state $\rho$ is given to us according to 
the unitary invariant distribution on the pure-state space, implying that 
we expect that each eigenvalue $\Omega_i$ occurs with equal probabilities 
as the outcome of the $\Omega$ measurement.  This {\it a priori} knowledge 
should be somehow taken into account in the estimation. 
We can see that this {\it a priori} knowledge is incorporated into the 
optimal estimate $\omega_{{\rm opt}}$ in a natural way. It is just the 
arithmetic average of $N$ observed data points $\{ \Omega_{i_n} \}_{n=1}^{N} $ 
and the $d$ ``default nonobserved'' data points $\{ \Omega_i \}_{i=1}^d$, 
the latter of which add up to the trace of the observable. 

One may still wonder why the weights of the average for the observed and 
non observed data are equal. Actually this is a feature of the pure-state 
ensemble considered in this paper. 
To see this, let us take the simplest example of $d=2$ and $N=1$, but this 
time the state $\rho$ is generally mixed. We assume that the Bloch vector 
$\bold{n}$ is distributed isotropically inside the Bloch sphere. 
The ensemble is characterized by the average $\langle n^2 \rangle$, 
which is 1 for the pure-state ensemble, but generally less than 1. 

After some calculation, the mean squared error turns out to be 
\begin{eqnarray}
 \langle \Delta \rangle &=& \frac{1}{2}\sum_a \tr{E_a 
           \left(\omega_a -\hat\Omega \right)^2}
                              \nonumber \\
   &+& \frac{\langle n^2 \rangle}{12}
      \left( 1-\frac{\langle n^2 \rangle^2}{3} \right) 
      \left( 2\tr{\Omega^2}-\left(\trm\Omega\right)^2 \right),
                     \label{Delta_mixed}
\end{eqnarray}
where
\begin{eqnarray}
   \hat\Omega = \frac{1}{3} \left(
      \frac{3-\langle n^2 \rangle}{2}\trm\Omega + 
              \langle n^2 \rangle \Omega \right).
\end{eqnarray}
This implies that the optimal measurement is the projective measurement 
of $\Omega$, and the optimal estimate is given by 
\begin{eqnarray}
   \omega_{{\rm opt}} = \frac{1}{3} \left(
      \frac{3-\langle n^2 \rangle}{2}\trm\Omega + 
              \langle n^2 \rangle \Omega_{i_1} \right),
\end{eqnarray}
where $\Omega_{i_1}$ is the observed eigenvalue of $\Omega$. 
The minimal mean squared error is given by the second term of 
Eq.~(\ref{Delta_mixed}).
We can see 
that the weight for the observed data decreases as the degree of mixing of the 
ensemble increases. 
When $\langle n^2 \rangle=0$, this $\omega_{{\rm opt}}$ implies we should 
disregard the observed data. The reason is that we know that the expectation 
value is given by $\trm\Omega/2$ for a completely mixed state.

The generalization of our analysis to an ensemble of mixed states, 
including the details of the above discussion, will be presented elsewhere.

\end{document}